\documentclass[manuscript]{emulateapj}
\usepackage[usenames,dvips]{color}
\usepackage{graphicx,natbib}
\usepackage{multirow}
\usepackage{amssymb}
\usepackage{ulem}

\shorttitle{HE emissions from Galactic BHTs}
\shortauthors{Lin et al.}

\begin{document}

\title{Searching For High Energy, Horizon-Scale Emissions from Galactic Black Hole Transients
       During Quiescence
     }

 \author{Lupin Chun-Che Lin${}^1$,
         Hung-Yi Pu${}^{2,1}$,         
         Kouichi Hirotani${}^1$, 
         Albert K. H Kong${}^{3,4}$,
         Satoki Matsushita${}^1$, 
         Hsiang-Kuang Chang${}^3$, 
         Makoto Inoue${}^1$, and
         Pak-Hin T. Tam${}^5$
         }
 \affil{${}^1$
        Institute of Astronomy and Astrophysics, Academia Sinica,
        Taipei 10617, Taiwan; \\
       lupin@asiaa.sinica.edu.tw; hpu@perimeterinstitute.ca; hirotani@tiara.sinica.edu.tw}
 \affil{${}^2$
       Perimeter Institute for Theoretical Physics,
       31 Caroline Street North, Waterloo, ON, N2L 2Y5, Canada} 
\affil{${}^3$
       Institute of Astronomy and Department of Physics, 
       National Tsing Hua University,
       Hsinchu 30013, Taiwan.}
 \affil{${}^4$
       Astrophysics, Department of Physics, University of Oxford, 
       Keble Road, Oxford OX1 3RH, UK}   
\affil{${}^5$
       School of Physics and Astronomy, Sun Yat-Sen University, 
       Zhuhai 519082, China}

\begin{abstract}
We search for the gamma-ray counterparts of stellar-mass black holes using long-term {\it Fermi} archive to investigate the electrostatic acceleration of electrons and positrons in the vicinity of the event horizon, by applying the pulsar outer-gap model to their magnetosphere. 
When a black hole transient (BHT) is in a low-hard or quiescent state, the radiatively inefficient accretion flow cannot emit enough MeV photons that are required to sustain the force-free magnetosphere in the polar funnel via two-photon collisions.
In this charge-starved gap region, an electric field arises along the magnetic field lines to accelerate electrons and positrons 
into ultra-relativistic energies.
These relativistic leptons emit copious gamma-rays via the curvature and inverse-Compton (IC) processes.
It is found that these gamma-ray emissions exhibit a flaring activity when the plasma accretion rate stays typically between 0.01 and 0.005 percent of the Eddington value for rapidly rotating, stellar-mass black holes.
By analyzing the detection limit determined from archival {\it Fermi}/LAT data, we find that the 7-year averaged duty cycle of such flaring activities should be less than $5$\% and $10$\% for XTE~J1118+480 and 1A~0620-00, respectively, and that the detection limit is comparable to the theoretical prediction for V404~Cyg.
It is predicted that the gap emission can be discriminated from 
the jet emission, if we investigate the high-energy spectral behaviour or 
observe nearby BHTs during deep quiescence simultaneously 
in infrared wavelength and very-high energies.

\end{abstract}

\keywords{black hole physics
       --- gamma rays: stars
       --- magnetic fields
       --- methods: analytical
       --- methods: numerical}

\section{Introduction}
\label{sec:intro}
In the past several years there has been increasing interest in the $\gamma$-ray emissions from the direct vicinity of accreting black holes (BHs).
Although accreting BHs can emit radiation in various wavelengths in general, only a few of them have the confirmed counterparts in the high-energy (HE) $\gamma$-ray band.
For example, Cyg X-3 (i.e., V1521 Cyg or 4U~2030+40) has a confirmed transient $\gamma$-ray detection using the {\it AGILE} data above 100 MeV \citep{Tavani2009} and the {\it Fermi}/LAT data between 100 MeV and 100 GeV \citep{Fermi2009}.
Cyg X-1 (i.e., V1357 Cyg or 4U~1956+350) also has a confirmed $\gamma$-ray counterpart detected by {\it Fermi}/LAT between the energy range of 60~MeV and 500~GeV \citep{Zanin2016}. 
On the other hand, several similar cases did not gain a positive detection in the $\gamma$-ray band \citep{Bodaghee2013}.
The observed $\gamma$-rays were concluded to be associated with the radio flares that are consistent with the radio flux level of the relativistic jets and shock formation in the accretion process.
Shocks propagating in a jet \citep[i.e., shock-in-jet model;][]{MG85,bjo00,tur11} provide a possible explanation of the (very) HE emission for Cyg X-3 \citep{MJ2009,Corbel2012} and for Cyg X-1 \citep{MZC2013}.

Recently, rapidly varying, sub-horizon scale TeV emissions were discovered from IC 310 \citep{Alek2014} using {\it MAGIC} (Major Atmospheric Gamma-ray Imaging Cherenkov) telescopes.
The observed radio jet power, the expected cloud crossing time and the proton-proton cooling time cannot support such models as jet-in-a-jet \citep{GUB2010} and clouds/jet interactions \citep{BP1997,BAB2010} for the shock-in-jet scenario.
A likely mechanism for such event-horizon-scale $\gamma$-radiation, is the particle acceleration taken place in the vacuum gap that arises in the polar funnel of a rotating BH magnetosphere \citep{bes92,hiro98,nero07,levi11,brod15,hiro16a}.

In additional to the emission site, there are several key differences between the shock-in-jet models and the BH gap models.
In the former, the spectrum is determined by the energy releases in the shock and the spectrum evolves as the shock propagating along the jet, with different phases (from Compton to synchrotron to adiabatic phases). A positive correlation between accretion rate and a shock emission is expected, provided that jet or outflow become more powerful at higher accretion rates \cite[however, observations of black hole X-ray binaries indicates a complicated disk-jet coupling pattern, see, e.g.,][]{fen04}.
In the latter, BH gap model, the rotational energy of a rotating BH is electrodynamically extracted via the Blandford-Znajek process \citep{bla77}, and dissipated in the form of charged-particle acceleration, which leads to a resultant $\gamma$-radiation from the direct vicinity of the BH.
Since the magnetic-field-aligned electric field is less efficiently screened
by the created pairs when the soft photon field is weak,
the $\gamma$-ray luminosity of a BH gap increases with decreasing mass accretion rate, 
$\dot{M}$ (\citealt[][hereafter H16]{hiro16b}).
What is more, although Cyg X-3 and Cyg X-1 are persistent BHs in high-mass X-ray binaries (HMXBs)  with high accretion rates, 
$\dot{M} \sim 3.64\times 10^{-8} M_{\odot}$~$\rm{yr}^{-1}$ and 
$\dot{M} \sim 3.88\times 10^{-9} M_{\odot}$~$\rm{yr}^{-1}$ 
(\citealt{Tet2016}; hereafter T16), 
and serve as examples of the shock-in-jet scenario,  
the BH-gap model provides alternative explanation of HE emissions
at lower accretion rate.
Furthermore, it is demonstrated in H16 that a BH gap can emit photons mostly in the HE (GeV) range via the curvature process for stellar-mass BHs, and mostly in the very-high-energy (VHE; TeV) range via the inverse-Compton (IC) process for super-massive BHs.

Motivated by the striking differences between the shock-in-jet and gap-emission scenarios, we expect that only BH transients (BHTs) in quiescence are plausible HE (GeV) gap emitters. 
As the first trial for seeking possible BH gap emission from stellar mass black holes, we select and concentrate on nearby BHTs with low accretion rates in low-mass X-ray binaries (LMXBs) and search for their HE emission by analyzing their 7-year archival data of the {\it Fermi}/Large Area Telescope (LAT).
In \S~\ref{sec:source}--\S~\ref{sec:obs}, we describe how we select the sources, estimated their mass accretion rate, $\dot{M}$, and analyzed their LAT archival data.
Then in \S~\ref{sec:model}, we outline the BH gap model.
In \S~\ref{sec:results}, we derive the observational upper limits on their HE fluxes and constrain their $\dot{M}$.
Finally in \S~\ref{sec:disc}, we compare the expected spectral behaviour obtained by the BH-gap and the shock-in-jet scenarios, and discuss the next targets to be observed in HE and VHE.
The distinct spectral features between the two models should help us discriminate the emission models with future HE observation.
Nevertheless, we also emphasise that even for a positive result of HE emission for the sources that have low accretion rates satisfying the requirement of BH gap models, the shock-in-jet model is not ruled out. 
The emission nature should be further determined by the overall spectral profile in the HE/VHE region, or/and its dependence on $\dot{M}$.

\section{Source selection}
\label{sec:source}
Using the recent BH gap model (H16),
we can infer which BHTs will exhibit strong HE and VHE fluxes at Earth.
Specifically, we can introduce the following conditions
to search for plausible gap emitters:\\
(1) The BH mass is large.\\
(2) The BH spin is large.\\
(3) The dimensionless accretion rate, $\dot{m}$, lies 
between $ $$10^{-4.25} $$\lesssim$$ \dot{m}$$\lesssim$$ 10^{-4}$$ $ near the horizon,
where $ $$\dot{m}$$ \equiv$$ \dot{M}/\dot{M}_{\rm Edd}$$ $, and
$\dot{M}_{\rm Edd} \equiv 2.18 \times 10^{-8} 
 (M/M_\odot) 
 (\eta/0.1)^{-1} 
 M_\odot \mbox{ yr}^{-1}
$
denotes the Eddington accretion rate with accretion efficiency 
$\eta \sim 10\%$.\\
(4) The distance is short. \\
(5) The observer's viewing angle, $\zeta_{\rm obs}$, 
is relatively small with respect to the rotation axis
(i.e. the binary system is nearly face-on; 
see \S~5.1.2 of H16 for details).\\

Conditions (1)--(3) represent intrinsic properties, 
while (4) and (5) show positional conditions relative to us.
In this letter, we define that a BHT is in \lq deep quiescence'
if condition (3) is met. 
In general, when the accretion rate is very low, 
the corresponding radiatively inefficient accretion flow (RIAF)
cannot supply enough MeV photons via free-free process to sustain the force-free magnetosphere
(e.g., fig.~1 of H16).
In this charge-starved magnetosphere, 
an electric field inevitably arises along the magnetic field line 
to accelerate electrons and positrons into ultra-relativistic energies.
The resultant gap emission becomes particularly strong when condition (3) is met.

It is noteworthy that persistent X-ray sources will not show 
strong gap emissions,
because their $\dot{m}$ is much higher than condition~(3).  
However, if a BHT has a lower mass-transfer rate
(e.g., $\dot M \lesssim 10^{-9} M_{\odot} \mbox{ yr}^{-1}$; 
 \citealt{TS96}), 
we can expect a stronger gap emission from such objects.
What is more, we can approximately estimate the gap luminosity 
from the BZ power (\S~2 of H16),
without solving the set of Maxwell-Boltzmann equations (\S~4 of H16),
if we impose a condition for the polar funnel to be highly charge-starved.
Utilizing this approximation, and using the BH masses and 
distances reported by T16, 
we select top four BH X-ray binaries whose BZ flux 
(i.e. BZ power divided by distance squared) is the greatest at Earth.
Among the four objects, we exclude the high-mass X-ray binary
Cyg~X-1, because its persistent X-ray emission indicates
$\dot{m}>10^{-4}$ in the entire period.
Coincidentally, the other three targets all have firm detections of jets, and some studies mentioned in \S~7
are interested in the connection between the jet detection and the HE emission.   
They are: (i) 1A 0620-00 (i.e. 3A 0620-003 or V616 Mon; \citealt{Kuuk99}), 
(ii) XTE~J1118+480 (i.e. KV~UMa; \citealt{BJMK2010}), and 
(iii) V404 Cyg (i.e. GS~2023+338; \citealt{GFH2005}). 
In this paper, we will examine these three objects as potential gap emitters.

\section{Detectability of BH gap emission}
\label{sec:mdot}
To infer the feasibility of strong gap emissions,
we must estimate the $\dot{m}$ 
for individual sources.
However, such an $\dot{m}$ is, in general, difficult to be constrained on real-time basis
unless the Faraday rotation is measured like the case of supermassive BHs \citep{Bower2003,Kuo2014}.
Nevertheless, their long-term averaged values can be 
estimated as described below.

For 1A~0620-00, the BH mass and the distance are inferred to be 
$M \approx 6.60 M_\odot$ and  $d \approx 1.06$~kpc, 
respectively, and our viewing angle is reported to be
$\zeta_{\rm obs} = 51.0\pm 0.9{}^\circ$ \citep{Cantrell2010}.
From the luminosity of the 1975 outbursts and the interval 
from the previous one in 1917, the mass accretion rate of 1A~0620-00 
can be estimated as $\dot{M}=3\times 10^{-11} M_{\odot}$~$\rm{yr}^{-1}$ 
\citep{MPRR83}, 
which corresponds to the dimensionless accretion rate 
$\dot{m} = 2.08 \times 10^{-4}$.
During quiescence, its accretion rate can be alternatively estimated 
by observing its bright spot, and can reach up to 
$\dot{M}=3.4\times 10^{-10} M_{\odot}$~$\rm{yr}^{-1}$ \citep{Fron2011}, 
or equivalently $\dot{m}= 2.36 \times 10^{-3}$, which is one-order higher than the previous measurement.
If $\dot{m}$ is close to the former value, $2 \times 10^{-4}$, 
which is slightly higher than condition~(3),
1A~0620-00 is expected to spend a certain fraction of time 
in the narrow $\dot{m}$ range defined by condition~(3), 
and hence to show HE and VHE flares.
However, if the actual $\dot{m}$ is close to the latter value, $2 \times 10^{-3}$, 
we do not expect that this source undergoes flaring activities
during a significant fraction of time. 
Its $\zeta_{\rm obs}$ indicates that a strong gap emission may be 
emitted marginally toward us (H16), 
if the polar funnel exists down to the lower latitudes
(e.g., to the colatitudes $80^\circ$)
within $6 r_{\rm g}$
\citep{mckinney12}.

\begin{table*}
\caption{{\small Selected BHTs to search for $\gamma$-ray emission from stationary black hole gaps.}}\label{BHTs}
\begin{tabular}{ccccccccc} 
\hline \hline {\footnotesize Source Name} & {\footnotesize RA (J2000)} & {\footnotesize DEC (J2000)} & {\footnotesize TS in} & {\footnotesize $F_{0.07-300 \rm{GeV}}^a$ } & {\footnotesize TS in} & {\footnotesize $F_{0.1-1 \rm{GeV}}^a$ } & {\footnotesize TS in} & {\footnotesize $F_{1-10 \rm{GeV}}^a$ } \\
 &  &  & {\scriptsize 0.07-300 GeV} & {\scriptsize (TeV/s/$\rm{cm}^2$)} & {\scriptsize 0.1-1 GeV} & {\scriptsize (TeV/s/$\rm{cm}^2$)} & {\scriptsize 1-10 GeV} & {\scriptsize (TeV/s/$\rm{cm}^2$)}
\\ \hline 
{\small 1A 0620-00} & {\small $06^h22^m44^s.5$} & {\small $-00^{\circ}20'44''.7$} & 1.91 & {\small $< 6.02 \times 10^{-13}$} & 0.00 & {\small $< 3.08 \times 10^{-13}$} & 0.00  & {\small $< 1.46 \times 10^{-13}$}
\\ {\small XTE J1118+480} & {\small $11^h18^m10^s.8$} & {\small $+48^{\circ}02'12''.4$} & 0.54 & {\small $< 5.58 \times 10^{-13}$} & 0.35 & {\small $< 3.56 \times 10^{-14}$} & 2.33 & {\small $< 9.20 \times 10^{-14}$}
\\ {\small V404 Cyg} & {\small $20^h24^m03^s.8$} & {\small $+33^{\circ}52'01''.9$} & 0.31 & {\small $< 3.86 \times 10^{-12}$}  & 1.64 & {\small $< 3.36 \times 10^{-13}$} & 0.40 & {\small $< 2.05 \times 10^{-13}$}
\\
\hline 
\multicolumn{9}{l}{{\footnotesize $^{a}$ The 2$\sigma$ flux upper limit was derived by the best fit with a single power-law to describe the spectrum.}} 
\\
\hline 
\hline
\end{tabular}

\end{table*}


For XTE~J1118+480, we have $M \approx 7.30 M_\odot$ (T16), 
$d \approx 1.72$~kpc \citep{Gelino2006}, 
and $68^\circ < \zeta_{\rm obs} < 79^\circ$ \citep{Kharg13}.
Its time-averaged mass accretion rate is estimated to be 
$\dot{M}=1.55\times 10^{-11} M_{\odot}$~$\rm{yr}^{-1}$ 
(updated table for 
 T16\footnote{http://142.244.87.173/static/MassTransferRate.txt?}), 
 or $\dot{m}= 9.92 \times 10^{-5}$.
This small $\dot{m}$ shows that the BH gap of XTE~J1118+480 will exhibit HE and VHE flares
in a significant fraction of time.
However, its large viewing angle indicates that the
BH-gap emission probably propagates away from our line of sight.

For V404~Cyg, we have $M \approx 7.15 M_\odot$ (T16), 
$d \approx 2.39$~kpc \citep{MJ2009}, and
$\zeta_{\rm obs}= 67^\circ{}^{+3}_{-1}$ \citep{Kharg10}.
Its time-averaged accretion rate is between 
$2.7\times 10^{-10}$ -- $3.5\times 10^{-9}$ $M_{\odot}$~$\rm{yr}^{-1}$ 
among its evolution track \citep{King93}, or 
$\dot{m}= 1.73 \times 10^{-3}$ -- 
 $2.24 \times 10^{-2}$.
But if we consider the updated measurement calculated between 
1996 January - 2016 September (updated table for T16$^1$), 
the time-averaged accretion rate is only 
$2.30\times 10^{-12}$ $M_{\odot}$~$\rm{yr}^{-1}$ 
corresponding to a much smaller $\dot{m}= 1.47 \times 10^{-5}$.
If the actual accretion rate is close to the former value,
it is very unlikely for V404~Cyg to exhibit a BH gap emission
during a large fraction of time.
However, if it is close to the latter value,
its BH gap will exhibit a strong HE and VHE emission,
which may be stationary or non-stationary,
with a large duty cycle.
However, its relatively large $\zeta_{\rm obs}$ may prevent us
from detecting these $\gamma$-rays,
which may propagate away from our line of sight.    

Although the $\zeta_{\rm obs}$'s of these target sources are large, 
we examine these three sources as the first step, because the actual $\zeta_{\rm obs}$ might be different from the represented values mentioned above.
In addition, BHTs in quiescence are known to be variable X-ray sources \citep{KMGMB2002}.
For instance, V404 Cyg can vary by a factor of 20 during quiescence 
\citep{KMGMB2002,Hynes2004}.
Thus, during a certain fraction of time, there is a possibility 
that their $\dot{m}$ enters this relatively narrow range to activate the gap.
Given that current observational evidence are not precise enough to determine the time interval for a deep quiescent stage\footnote{In comparison, HE emission is preferentially expected to be accompany with a prominent jet detection or at the outburst stage in other traditional models \citep{MG85}.},
as the first trial, we analyze the archival LAT data and examine their time-averaged HE fluxes during August 2008 - November 2015.

\section{Observations and data analysis}
\label{sec:obs}
LAT is a wide-band $\gamma$-ray detector (in 20 MeV -- 300 GeV) 
onboard the {\it Fermi} observatory, which provides all-sky survey 
every two orbits since August 2008.
Until the end of 2015, we already have 7-year accumulation time for a long-term investigation for those stellar-mass BHTs.
For all the targets in our interests summarized in \S 1, we downloaded the Pass 8 (P8R2) archive from the LAT data server\footnote{http://fermi.gsfc.nasa.gov/cgi-bin/ssc/LAT/LATDataQuery.- \\ cgi} within a circular FOV in $10^{\circ}$ radius and considered the time distributed among August 2008 to November 2015 for further investigations.
For V404 Cyg, which experienced two outbursts in 2015, the photons detected during the first outburst (June/July; \citealt{Jenke2016}) are removed from the analysis and our data do not cover the epoch of the second outburst \citep{Motta2015}, so that we collect only the photons during low accretion rates.
  
In order to reduce and analyze data, we used the {\it Fermi} Science tools v10r0p5 package and constrain photons in the class for the point source or galactic diffuse analysis (i.e. evclass = 128). 
We also collected all the events converting both in the front- and back-section of the tracker (i.e. evtype = 3), and the corresponding instrument response described for the selected event type has been defined in the ``P8R2\_SOURCE\_V6'' IRF (Instrument Response Function) used throughout this study. 
Events with zenith angles larger than $90^{\circ}$ were also excluded to avoid the contamination from Earth albedo $\gamma$-rays, and only the good data quality was counted in the time selection to exclude those data within time intervals affected by some spacecraft events (i.e. DATA\_QUAL$>0$).   

We performed the binned likelihood analysis using the ``NewMinunit'' optimization algorithm by defining a $7^{\circ} \times 7^{\circ}$ square region of interest (ROI), together with the long time span and the energy range of 0.07-300 GeV.
The choice for the lower boundary of the energy range was constrained by the usage of IRF to generate the exposure map, and we included the energy dispersion correction when the analysis includes photons $<$100~MeV.
The source model determined for likelihood analysis is based on the LAT 4-year point source (3FGL; \citealt{Acero2015}) catalog, and the spectral parameters of each source in ROI is freed to gain the best fit to the archival observation. 
The standard templates of Galactic and isotropic background (gll\_iem\_v06.fits \& iso\_P8R2\_SOURCE\_V6.txt) are included in our analysis as well.
The central positions determined for our targets are summarized in Table~\ref{BHTs}, and they can be referred to  
\citet{Gallo2006,Bailyn95,MJ2011} and \citet{MJ2009}. 
Since all of our targets are not known {\it Fermi} sources, a simple power law was adopted to characterize their spectra throughout the analysis.
The test-statistic (TS) values yielded with the best fit to observations of the different energy range are also reported in Table~\ref{BHTs}.   

\begin{figure}[t]
\includegraphics[angle=0,scale=0.50]{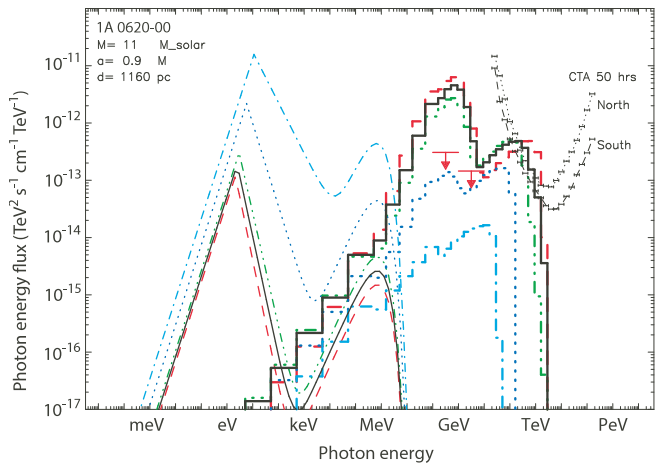} 
\caption{\footnotesize{
Spectral energy distribution (SED) of the gap emission from 1A~0620-00 
with BH mass of $M=11 M_\odot$, dimensionless spin of $a_\ast=0.9$, and 
distance of $d=1.16$~kpc.
The lines correspond to different dimensionless accretion rates, 
$\dot{m}=1.00 \times 10^{-3}$ (cyan dash-dotted) , $10^{-3.5}$ (blue dotted), $10^{-4}$ (green dash-dot-dot-dotted), $10^{-4.125}$ (black solid), and $10^{-4.25}$ (red dashed), respectively.
A stationary vacuum gap is expected to take place when $10^{-4.25}<\dot{m}<10^{-4}$.
The thin curves on the left denote the input ADAF spectra, 
while the thick lines on the right do the output gap spectra.
The red down arrow shows the upper limit flux obtained by 
re-analyzing the {\it Fermi}/LAT archival data.
The thin dashed and dotted curves (with horizontal bars) 
denote the CTA detection limits after a 50-hour observation.
Magnetic field strength is assumed to be the equipartition value 
with the plasma accretion. 
}}
\label{fig:SED_0620}
\end{figure}


\section{The black hole gap model}
\label{sec:model}
To compare the LAT observational constraints with the theoretical prediction, we apply the method described in \S~4 of H16 to individual BHTs.
Namely, we solve the set of the following three differential equations.
First, we solve the {\it Poisson equation} for the non-corotational potential (eq.~[19] in H16) near the event horizon. 
The magnetic-field-aligned electric field, $E_\parallel$, can be computed from the Poisson equation through eq.~(23) of H16.
Second, we solve the {\it Boltzmann equations} of the produced $e^\pm$'s in the gap, assuming that their Lorentz factors saturate at the curvature-limited terminal value at each position.
This assumption is valid for stellar-mass BHs, because the curvature process dominates the IC one, and because the acceleration length is shorter than the gap width particularly during the HE flare.
Third, we solved the {\it radiative transfer equation} of the emitted photons, assuming they have vanishing angular momenta.
 
We adopt the analytic solution of \citet{Mahadevan97} to describe the RIAF soft photon, and solve the gap in the 2-D poloidal plane (\S~4.2.5 of H16).
Both the outer and inner boundary positions were solved from the free boundary problem.
To estimate the greatest gap flux, we adopt $a_\ast=0.9$ and 
$\Omega_{\rm F}=0.5\omega_{\rm H}$, where $a_\ast \equiv a/r_{g}$ is the dimensionless BH's spin parameter, and $r_{\rm g} \equiv GMc^{-2}$ represents the gravitational radius; $\Omega_{\rm F}$ and $\omega_{\rm H}$ denote the angular frequency of rotating magnetic field lines and a rotating BH, respectively.
We assume that the magnetic field takes the equipartition value with the plasma accretion,
\begin{equation}
  B=B_{\rm eq} 
    \approx 4 \times 10^8 \dot{m}^{1/2} M_1{}^{-1/2} 
            \mbox{ G},
  \label{eq:B_eq}
\end{equation}
at $r= 2 r_{\rm g}$, where $M_1$ denotes the BH mass in ten solar-mass unit.

\section{Results}
\label{sec:results}
 
In spite of the long-term accumulation of the {\it Fermi} data, we did not detect any counterparts at a significant level for the three sources we have concerned.  
This result is in consistency with sources resolved in 4-year {\it Fermi} catalogue \citep{Acero2015} and the examination for V404 Cyg using 7-year LAT data \citep{LCD2016}.  
We also tried the energy-resolved investigations as well; however, the corresponding $\gamma$-rays emitted from the direction of the source position are still too few to show any clear signature in the binned likelihood analysis.
Assuming a simple power-law spectrum, we find that the spectral parameters 
have very large uncertainties in this case.

\begin{figure}[t]
\includegraphics[angle=0,scale=0.50]{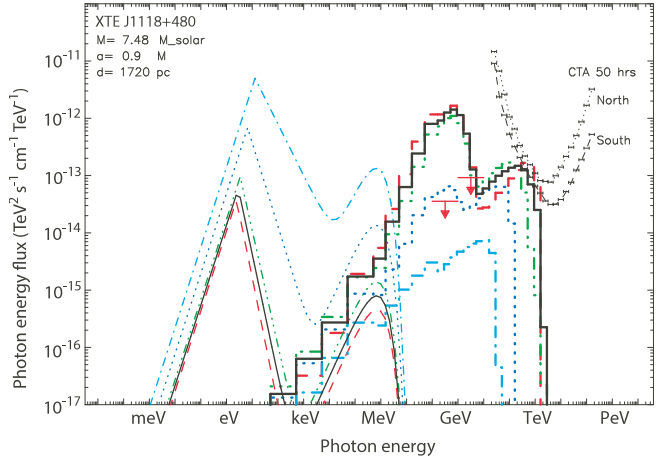}
\caption{\footnotesize{
Same as Fig.~\ref{fig:SED_0620} but for XTE~J1118+480,
with $M=7.48 M_\odot$, $a_\ast=0.9$, and $d=1.72$~kpc.
The lines for the gap model correspond to the same $\dot{m}$ as Fig.~\ref{fig:SED_0620}.
}}
\label{fig:SED_J1118}
\end{figure}

\begin{figure}[t]
\includegraphics[angle=0,scale=0.50]{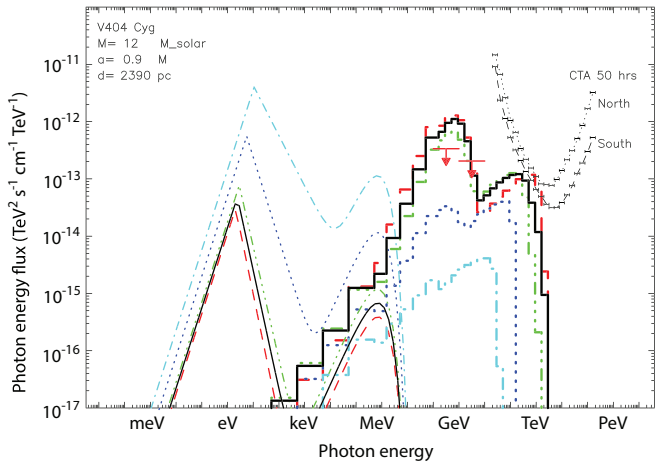}
\caption{\footnotesize{
Same as Fig.~\ref{fig:SED_0620} but for V404~Cyg,
with $M=12 M_\odot$, $a_\ast=0.9$, and $d=2.39$~kpc.
The lines for the gap model correspond to the same $\dot{m}$ as Fig.~\ref{fig:SED_0620}.
}}
\label{fig:SED_V404Cyg}
\end{figure}

On these grounds, we list their $2\sigma$ flux upper limits in Table~\ref{BHTs}, according to the best spectral fit obtained in the analysis.
It is also noteworthy that the VHE flux may be detectable below 1~TeV during a HE flare, as indicated by the sensitivity curves of CTA in Figs.~\ref{fig:SED_0620}--\ref{fig:SED_V404Cyg} (dashed and dotted curves labelled with \lq\lq CTA 50 hrs'')\footnote{https://www.cta-observatory.org/science/cta-performance/- \\ \#1472563157332-1ef9e83d-426c}.
In all the figures, we put these upper limits on the predicted spectra of their BH-gap emissions.

Fig.~\ref{fig:SED_0620} shows the case of 1A~0620-00.
The cyan dash-dotted, blue dotted, green dash-dot-dot-dotted, black solid, and red dashed lines correspond to the dimensionless 
accretion rates, $\dot{m}=1.00 \times 10^{-3}$ , $10^{-3.5}$, $10^{-4}$, $10^{-4.125}$, and $10^{-4.25}$, respectively.
The latter three lines show that the flux flares up in 0.1--3~GeV via curvature process and in 0.03--1~TeV via IC process, when the dimensionless accretion rate is between $5 \times 10^{-5}$ and $10^{-4}$.
We then put the flux upper limits (obtained from the 7-year LAT archival data as described above) with the two red down arrows in 0.1--1~GeV and 1--10~GeV.
It follows that the flaring HE fluxes,
which are represented by the green, black, and red lines, 
are predicted to be 10--20 times greater than the observational upper limits.
Thus, we find that the 7-year averaged duty cycle of such flaring activities was less than $10$\% from August 2008 to November 2015 for 1A~0620-00, provided that $a_\ast > 0.9$ and $\zeta_{\rm obs} < 40^\circ$ 
(H16), and adopting a conservative flux estimate at $\dot{m}=10^{-4}$.
For instance, if the BH gap flares during $10$\% of the entire period,
its flux is predicted to appear at (or slightly above)
the $10$\% level of its peak, 
which is comparable with the observational upper limit
(red down arrow).
However, if its BH is rotating as slowly as $a_\ast \sim 0.12$ \citep{bambi16}, there is no chance for such a slow rotator to emit a detectable flux at Earth from its gap; that is, we will not be able to obtain any constraints on the duty cycle of HE flares, using the BH gap model.

In the same way, we also apply the BH gap model to the two other targets and generate the predicted spectra (Fig.~\ref{fig:SED_J1118} \& Fig.~\ref{fig:SED_V404Cyg}). 
Comparing with the observational upper limits (red down arrows) obtained by the binned likelihood analysis with long-term {\it Fermi} archive, we find that the duty cycle of the BH gap is less than 
$0.05$ (i.e.\,5\,\%)
for XTE~J1118+480, provided that $a_\ast \ge 0.9$ 
and $\zeta_{\rm obs} < 40^\circ$. 
For V404~Cyg, the predicted HE flux lies at the same level of the observational upper limits.
Thus, we cannot constrain the duty cycle of its gap flares.

Lastly, we would like to comment on the dependence between $a_\ast$ and BZ power.
The BZ power, and hence the gap luminosity is approximately proportional to $a_\ast{}^2$.
(It is noteworthy that the BH gap emission and the jet HE emission are independent each other.)
Thus, the predicted flux is approximately halved
if $a_\ast$ reduces from 0.90 to 0.45, for instance.
In the case of a smaller $a_\ast$, 
the upper limit of the gap flaring duty cycle will be less constrained.
On the other hand, if $a_\ast>0.95$ is observationally confirmed,
it is no longer valid to assume 
a constant radial magnetic field,
$B^r \propto F_{\theta\varphi}/\sqrt{-g}$, on the horizon,
where $F_{\theta\varphi}$ denotes meridional derivative of 
the magnetic flux function,
and $\sqrt{-g}$ the volume element
\citep{tanabe08,tchekhov10}.
Such an extremely rotating case will be investigated in a separate paper.

\section{Discussion}
\label{sec:disc}
The non-detection of the HE fluxes from any of the three sources, may indicate either that their long-term accretion rates are above condition~(3) (i.e. $\dot{m} > 10^{-4}$, see \S~\ref{sec:source}) so the duty cycle of the gap activity is weak, or that our line of sight misses their $\gamma$-ray flares.
It is also possible that the BH spin is actually less than what we assumed.
For example, even for 1A~0620-00 or XTE~J1118+480,
it is very unlikely to detect their BH-gap emission with LAT
even if its flaring duty cycle is nearly 100~\%,
if $a_\ast < 0.25$.
Nevertheless, in general, there might be a higher possibility for a BH binary to show detectable gap emission if its time-averaged accretion rate is moderately small, and if we view the binary almost face-on.
We can expect a small accretion rate for the sources that stay in long quiescence and their recurrence time is long.
For instance, if we simply assume a BH mass of 10~$M_\odot$, XTE~J1818-245 is estimated to have relatively close accretion rate to condition~(3) (T16), although it does not have confirmed $M$, or $\zeta_{\rm obs}$ yet.
Such a source is located within several kpc (e.g., in the near side of our Galaxy), and its BH gap emission may be detectable during its deep quiescence if we luckily view them almost face-on (condition~5).
Since the accretion rate is variable, 
it is desirable to observe the aforementioned BHTs frequently
during their quiescence in near-IR and/or VHE, 
in order not to miss their gap flares.
For example, once the near-IR flux decreases enough,
we suggest to observe the source with ground-based, 
Imaging Atmospheric Cherenkov Telescopes (IACTs)
to detect their BH-gap emission below a few TeV.
In this case, X-ray flux is predicted to be very weak.

On the contrary, if we detect a HE flare contemporaneously with an X-ray flare (i.e., during its high accretion rate phase), it suggests that these photons are emitted from the jet, rather than the BH gap.
Thus, contemporaneous observations at X-ray and HE ranges will help us discriminate the emission processes in accreting BH systems.
One similar example is the investigation of V404 Cyg during its 2015 X-ray outburst in June using the {\it Swift}/BAT, INTEGRAL/ISGRI and {\it Fermi}/LAT data \citep{LCD2016}.
Even with the consideration of local time bin (i.e., 6-hours), the detection significance yielded from the unbinned likelihood analysis 
for the $\gamma$-ray data is still less than 4$\sigma$.
\citet{Piano2017} considered the {\it AGILE} data in the 50--400 MeV energy band and improves the significance of the source detection to $\sim 4.3\sigma$.
Inverse Compton scattering of photons applied to explain the transient $\gamma$-rays detected for Cyg X-3 \citep{Acero2015} can provide a similar scenario to support for the simultaneous detection of radio outburst, pair annihilation and HE $\gamma$-rays for V404 Cyg. 
However, the observed strong cut-off in the HE emission of $\sim$400 MeV may give a constraint to classify such a detection. 
Whether there exists a similar spectral component for the hard X-ray and $\gamma$-ray can also serve as another basis to definitely confirm the emission mechanism to the traditional shock-in-jet approach.      

Due to the non-detection of the HE emission or the correlation between the accretion rate and the HE flux, we cannot specify whether the shock-in-jet or the BH gap model is responsible for these sources.
To gain more insights into the differences between the gap model and the shock-in-jet model, we can compare the spectral features of both models.  
In the gap of a stellar-mass black hole, essentially all the electrons have the same Lorentz factor at each position, because their motion is saturated by the curvature radiation drags. 
On the contrary, in the shock-in-jet scenario, the electron is heated at the shock, resulting in a wider energy distribution than those electromagnetically accelerated in the gap. There exist 
a characteristic slope for the shock-in-jet model of the optically thin flux density, $\nu S_{\nu}\propto \nu^{-s/2+1}$, during the so-called ``Compton stage"  \citep[][]{MG85}.
During this stage, the energy density of the magnetic field is much larger than the energy density of photons when the shock propagates along the jet. 
With $s\approx2.1$ \citep[][]{SS11}, a characteristic slope of $\nu S_{\nu}\propto \nu^{-0.05}$ is estimated.
The  range of the such slope is related to the electron energy distribution, and the normalization is determined by the strength of the shock.
It is, therefore, possible to distinguish the responsible emission models according to the high energy spectral profile.
Namely, in the gap model, a bumpy spectrum is expected due to the nearly mono-energetic distribution of electrons.
On the other hand, in the shock-in-jet model, a smoother, singe power-law is expected due to a much wider, power-law energy distribution of electrons.
With current result of non detection, there is no observational preference for either model.

Moreover, inclusion of lower energy observations are also useful in discriminating the emission models.
It follows from Figs.~\ref{fig:SED_0620}--\ref{fig:SED_V404Cyg} that the gap HE and VHE fluxes increases with decreasing $\dot{m}$.
That is, we can predict an {\it anti-correlation} between the IR/optical and HE/VHE fluxes (H16).
It forms a contrast to the standard shock-in-jet scenario, in which the IR/optical and the HE/VHE fluxes will correlate.
If their time-varying multi-wavelength spectra show anti-correlation, it strongly suggests that the photons are emitted from the BH gap.
With the observational upper limits of the individual sources and the upper limits predicted by the BH-gap model, we can constrain their mass accretion rate, $\dot{M}$, which are potentially important to discuss their binary evolution through the long-term accretion rate.
Therefore, we propose to simultaneously observe nearby low-mass X-ray binaries that have more massive BHs during deep quiescence in near-IR/optical, X-ray, HE, and VHE (with CTA) in the future.

\acknowledgments
One of the authors (K. H.) is indebted to Drs. A. Okumura and T.~Y. Saito for a valuable discussion on the CTA sensitivity.
This work made use of data supplied by the LAT data server of Fermi Science Support Center (FSSC), and of the computational facilities in Theoretical Institute for Advanced Research in Astrophysics (TIARA) in ASIAA.
This work is supported by the Ministry of Science and Technology of the Republic of China (Taiwan) through grants 103-2628-M-007-003-MY3 and 105-2112-M-007-033-MY2 for A.~K.~H.~K., through grant 105-2112-M-007-002 - for H.~K.~C., through grant 103-2112-M-001-032-MY3 for S.~M., and
P.~H.~T.~T. is supported by National Science Foundation of China (NSFC) through grants 11633007 and 11661161010.

{\it Facilities:} \facility{Fermi (LAT)}. 


\end{document}